%% file: sample-acmtog-SIGGRAPH-submission.tex
\documentclass[acmtog]{acmart}

\renewcommand\footnotetextcopyrightpermission[1]{}

\usepackage{booktabs}
\usepackage{xcolor}
\usepackage[linesnumbered,ruled,vlined]{algorithm2e}
\usepackage{wrapfig}

\SetAlFnt{\small}
\SetAlCapFnt{\small}
\SetAlCapNameFnt{\small}
\SetAlCapHSkip{0pt}

\begin{document}

\title{Escher Tile Deformation via Closed-Form Solution}

\author{Crane He Chen}
\affiliation{
  \institution{Industrial Light \& Magic}
  \city{San Francisco}
  \state{CA}
  \country{USA}
}
\email{cranec@ilm.com}

\author{Vladimir G. Kim}
\affiliation{
  \institution{Adobe Research}
  \city{Seattle}
  \state{WA}
  \country{USA}
}
\email{vokim@adobe.com}

\begin{abstract}
We present a real-time deformation method for Escher tiles\textemdash interlocking organic forms that seamlessly tessellate the plane following symmetry rules. We formulate the problem as determining a periodic displacement field. The goal is to deform Escher tiles without introducing gaps or overlaps. The resulting displacement field is obtained in closed form by an analytical solution. Our method processes tiles of 17 wallpaper groups across various representations such as images and meshes. Rather than treating tiles as mere boundaries, we consider them as textured shapes, ensuring that both the boundary and interior deform simultaneously. To enable fine-grained artistic input, our interactive tool features a user-controllable adaptive fall-off parameter, allowing precise adjustment of locality and supporting deformations with meaningful semantic control. We demonstrate the effectiveness of our method through various examples, including photo editing and shape sculpting, showing its use in applications such as fabrication and animation.
\end{abstract}


\begin{teaserfigure}
\centering
\includegraphics[width=1.0\textwidth]{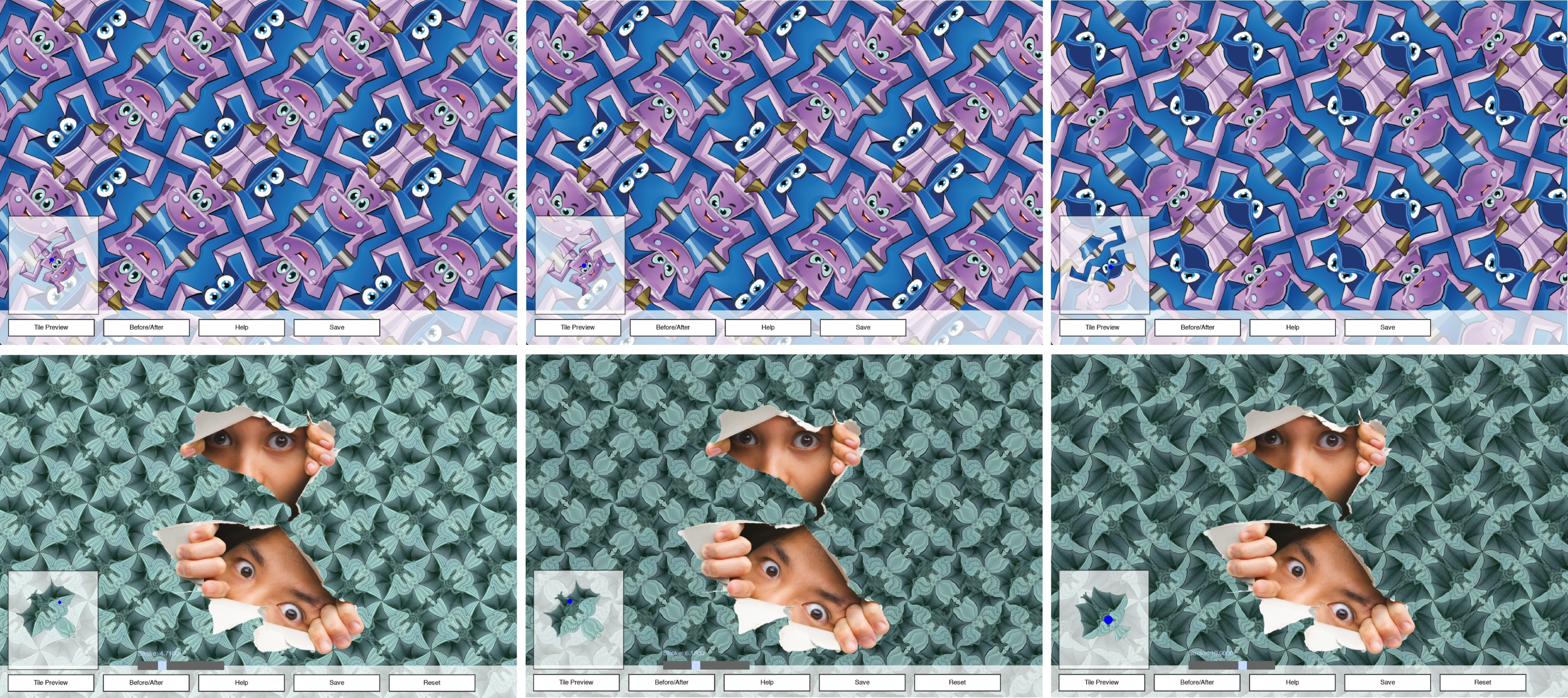}
  \caption{\textbf{Deforming both the boundary and interior of tiles via closed-form solution:} Examples of the proposed real-time interactive deformation method. Tileability is preserved before and after deformation. Second row: All M.C. Escher works © 2025 The M.C. Escher Company - the Netherlands. All rights reserved. Used by permission. www.mcescher.com }
  \label{fig: lamps}
\end{teaserfigure}

\maketitle

\input{samplebody-journals}

\end{document}

%% file: samplebody-journals.tex
\section{Introduction}
Tiles and patterns have been used in fabrication and architecture~\cite{jones_alhambra, jones_grammar_1856} across cultures since the early stage of civilization. One form of pattern commonly used is a tessellation: an infinite covering of an entire plane without gaps or overlaps by repeating a single artist-designed tile. These shapes can range from abstract geometries to representations of plants or animals. Observing the design of ornamental tiles~\cite{EscherBook}, the appeal lies not only in the interlocking shapes, but also in the rich textures. Thus, instead of representing tiles by their boundaries, we depict them as textured shapes.

\begin{figure}
\includegraphics[width=0.5\textwidth]{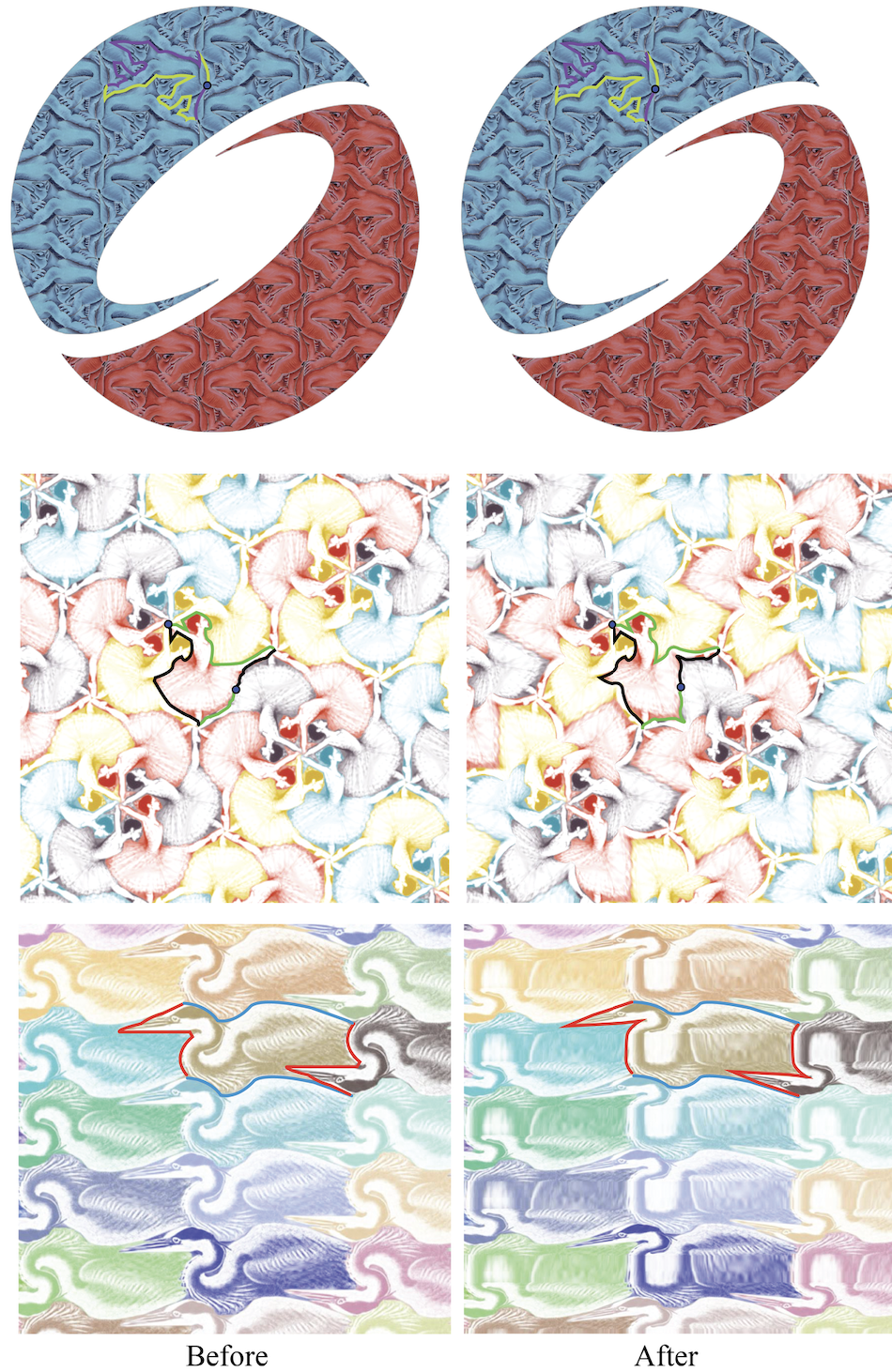}
  \caption{\textbf{Deformation with symmetry constraints:} A monster with muscular back. A ballet dancer bending knee. A heron straightening neck. First row: All M.C. Escher works © 2025 The M.C. Escher Company - the Netherlands. All rights reserved. Used by permission. www.mcescher.com }
  \label{fig: deform_generative_art}
\end{figure}

\begin{figure}
\includegraphics[width=0.5\textwidth]{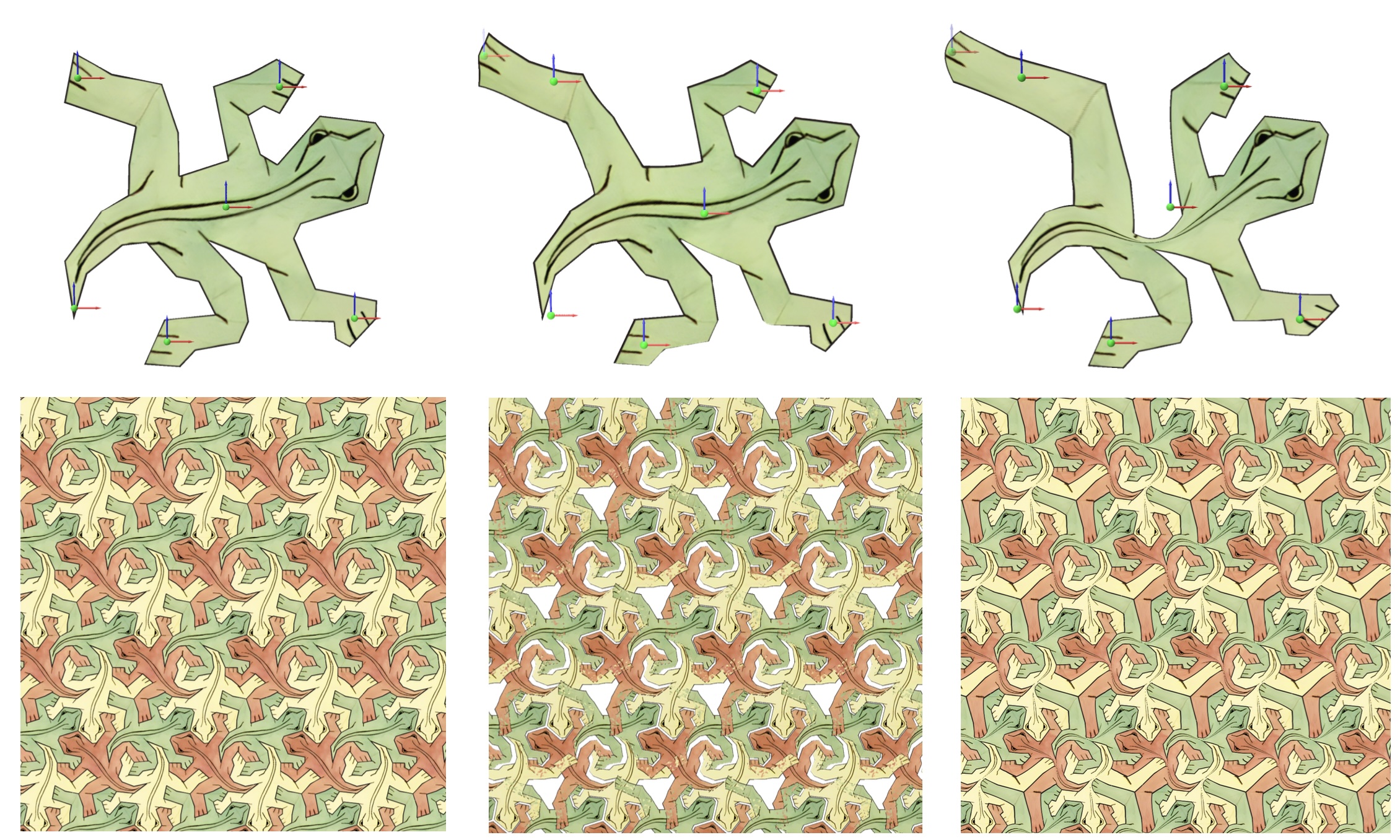}
  \caption{\textbf{Comparison with naive point handle-based deformation:} Editing the iconic gecko tile with the proposed tool. First column: before interactive editing. Second column: ARAP after editing. Third column: the proposed method after editing. All M.C. Escher works © 2025 The M.C. Escher Company - the Netherlands. All rights reserved. Used by permission. www.mcescher.com }
  \label{fig: gecko_teaser}
\end{figure}

Our primary focus is on the interactive deformation of tiles (Fig.~\ref{fig: deform_generative_art}). While tile editing tools often focus on boundary manipulation, we present a novel tile deformation approach that simultaneously adjusts the boundary and the interior texture. This unified method allows the artist to immediately see the changes to the tile's full appearance. In computer graphics, point handle based approaches are often used for interactive deformation. Users select and manipulate a set of points called handles. The deformation algorithm then automatically adjusts the entire object accordingly. However, existing interactive deformation algorithms cannot be used to deform tiles while maintaining tileability, as they are not designed to handle periodicity (second column of Fig.~\ref{fig: gecko_teaser}), easily leading to gaps and overlaps. The primary challenge posed by periodicity is that even if the user places just one handle, the infinite repetition of the tile causes the input handle to repeat infinitely as well. Each copy of the tile can be affected by more than one copy of the same handle who are not guaranteed to move in the same direction if there is a rotational or reflective transformation. At first sight,  the problem is not even well-defined. 

\begin{figure*}[thb]
\includegraphics[width=1.0\textwidth]{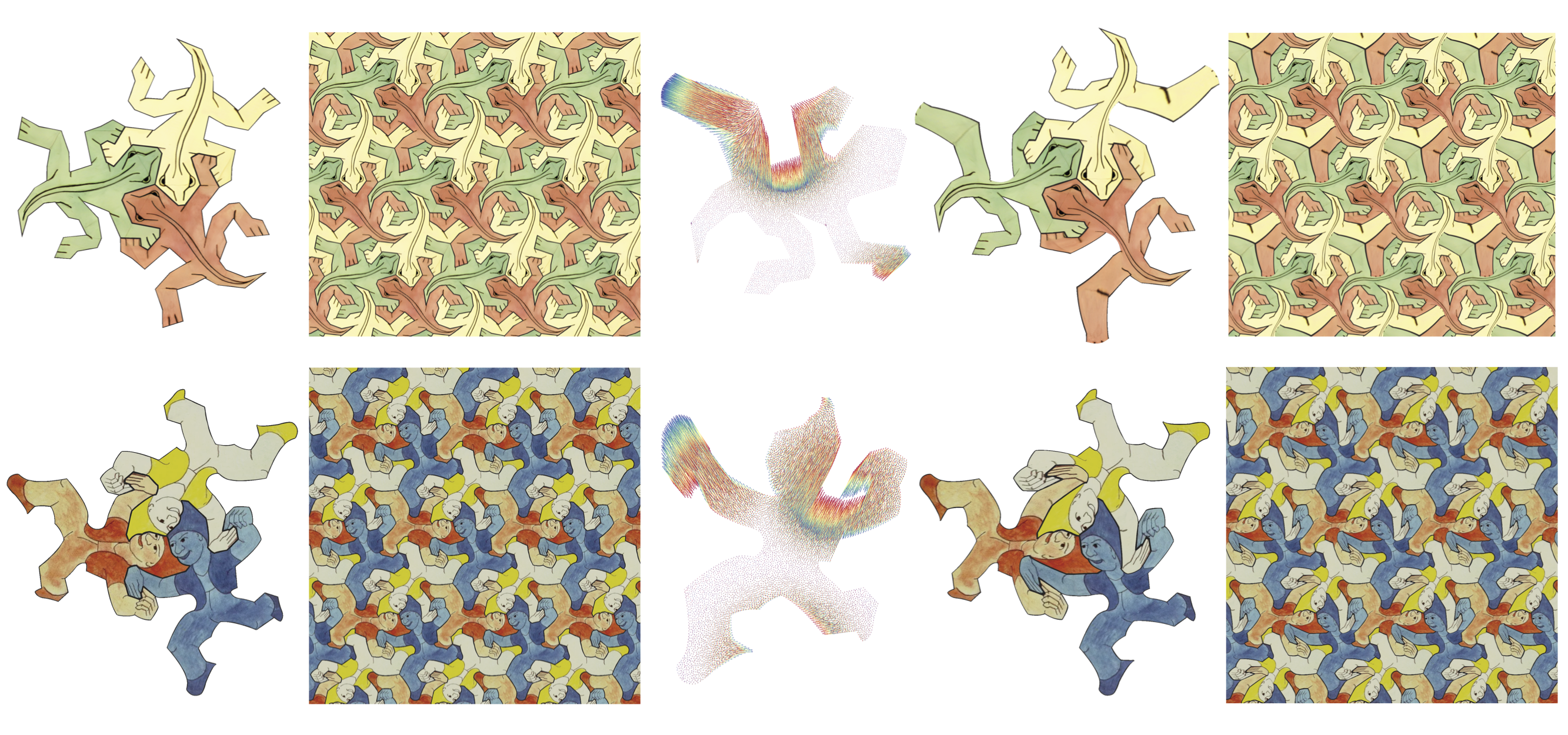}
  \caption{\textbf{``Infinite Clock":} The boundary of a tile is like an infinite clock. To remain a valid tile (that is the ability to cover the entire plane with no gap or overlap), when one boundary is extruding, there must be another boundary intruding. As can be observed from the displacement field, leg extruding and belly intruding of the gecko need to happen the same time, arm extruding and shoulder intruding of the clown need to happen the same time. All M.C. Escher works © 2025 The M.C. Escher Company - the Netherlands. All rights reserved. Used by permission. www.mcescher.com }
  \label{fig: infinite clock}
\end{figure*}
In this work, we present an analytical method to deform tiles with point handles. Key to the closed-form solution is decoupling translation from the other transformations, \textit{i.e.} rotation, reflection, and glide-reflection. A basic tile in a periodic tessellation can be interpreted as an orbifold~\cite{thurston1978geometry}, that is a coned manifold. The cone points of an orbifold (detailed in Section~\ref{chap:background}) and mirror lines transform the displacement of one handle (a vector) into a set of vectors that we call the \textit{fundamental vectors}. Then the pure translation duplicates each fundamental vector into a vector field that we call the \textit{guidance field}. The guidance field is simple to compute, as it consists of vectors with the same magnitude and direction, therefore, they are still one vector (the fundamental vector), just living at different vertices of a regular grid. Since the guidance field consists of parallel vectors, integrating its effect at an arbitrary location requires only computing the magnitude. Another perspective is that we are computing the contribution,~\textit{i.e.} weight, of the guidance field at the queried location. Since the problem reduces to calculating the weight of each fundamental vector, we derive three analytical solutions that comprehensively represent all 17 wallpaper groups on the Euclidean plane, covering all periodic tiling patterns. Observing that the displacement field needs to be a periodic function for the deformed shape to remain a valid tile, we calculate the closed-form solution within one cycle. Mathematical series is employed to integrate the guidance field analytically.

In summary, we propose a tool that deforms tiles in real time while maintaining tileability by solving a displacement field. Tiles are commonly represented as images, curves, or meshes. Since our method operates on the continuum, it is indifferent of the specific representation of the tiles. The tool can be used to to edit patterns in photos and illustrations or stand-alone tiles in 2D meshes or 3D volumes.

\section{Related Work}
 \paragraph{Tiling Theory in Computer Graphics:} Tiling theory~\cite{grunbaum1987tilings} has been studied by the graphics community for half a century~\cite{alexander1975computer, kaplan2009introductory}. This line of research was motivated by using the computer to understand the rich cultural heritage of architectural ornament, and to develop software tools that aid the design. They can be classified into abstract geometry~\cite{kaplan2004islamic}, plant forms~\cite{wong1998computer}, and animal forms~\cite{kaplan2000escherization}. While initially formulated for 2D plane, tiling theory also generalizes to 3D volumes~\cite{guerten2020kuhkubus}. It has the potential to be applied in designing efficiently packable hardware~\cite{chen2022computational,cui2023dense,goertzen2024constructing} such as self-replicating robots~\cite{chirikjian2002self}. From the perspective of math and art, tiles could be fascinating partly due to its linkage with infinity, chaos, and symmetry. In this paper, we focus on deforming symmetric tiles that regularly subdivide the Euclidean plane.

M.C. Escher is a pioneer mathematical artist renowned for tiles. In the pursuit of what interesting shapes might be tileable — M.C. Escher stands as a master. Although he did not seek closed-form solutions on lattices as we do, he investigated a special class of hexagons that tessellate the plane and proposed Escher’s theorem~\cite{rigby1991napoleon}, which states that such hexagons—known as Brianchon hexagons—must have congruent diagonals. Our deformation method operates in a way that maintains the congruent edges and congruent diagonals, aligning with his findings. Escher's artworks have inspired extensive research in the graphics community. Some studies explore fractals~\cite{ouyang2014beautiful, ouyang2021self, ouyang2022interlocking, schor2024into}, allowing tile scale variation, while others focus on fixed-area tiles~\cite{kaplan2000escherization, Kaplan2004DihedralE, lin2017generation, nagata2021escherization, aigerman2023generative}. Escherization~\cite{kaplan2000escherization} finds the closest valid tile to a given shape, while Generative Escher Meshes~\cite{aigerman2023generative} deforms a square mesh to match a text prompt. These works do not allow interactive deformation with explicit artist control (e.g., direct manipulation of handles). Other tile deformation methods~\cite{kaplan2008metamorphosis, kaplan2010curve, lee2015algorithms} focus on morphing between two target shapes, solving more of the interpolation problem than interactive editing. Since our method focuses on direct manipulation of tiles using point handles, we consider these deformation methods as complementary to ours.

\begin{figure}[thb]
\includegraphics[width=0.5\textwidth]{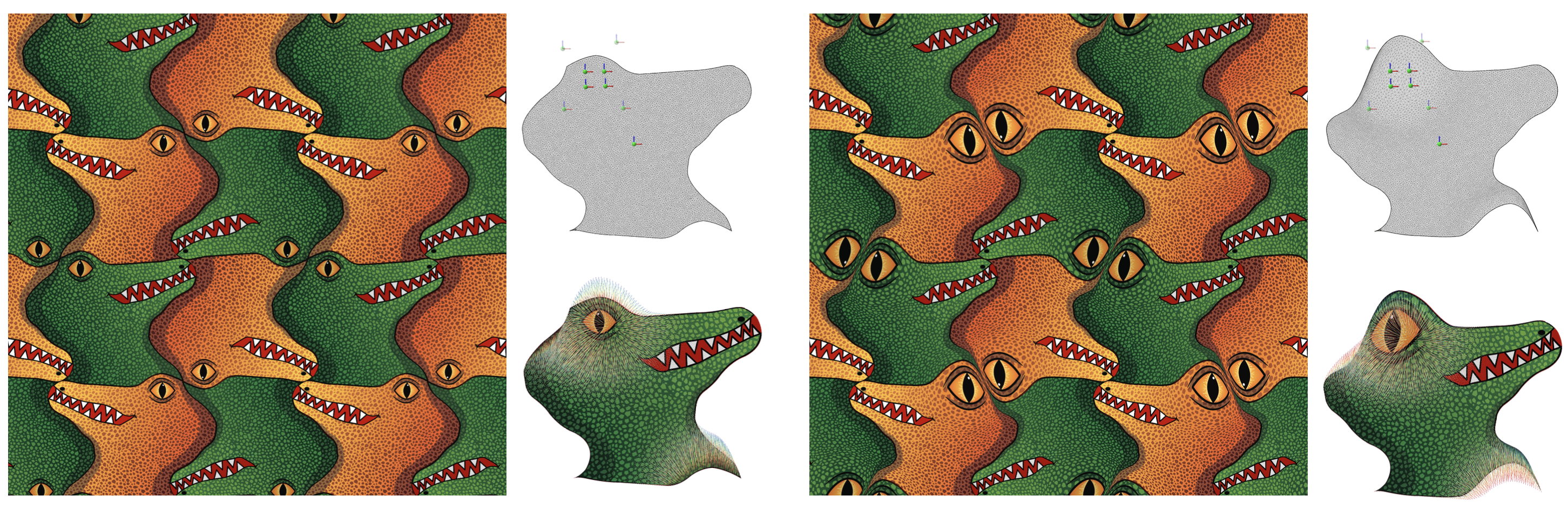}
  \caption{\textbf{Editing a tile represented by a 2D triangle mesh}: Enlarging the eye of the dinosaur head tile using multiple point handles. Left: before deformation. Right: after deformation. Input artwork adapted from \textcopyright{} mrhighsky from Adobe Stock.}
  \label{fig: eye_enlarging}
\end{figure}

Several interactive tools facilitate tile design. Conway's Magic Pen~\cite{conways_magic_pen} allows users to create tiles based on Conway’s Criterion~\cite{schattschneider1980will}, which requires a six-edged tile to have two congruent parallel edges and four centrosymmetric edges. However, it only supports 10 of the 17 wallpaper groups, specifically those with 180-degree rotational symmetry. KaleidoTile~\cite{kaleidotile} generates tiling from user-input images. Three images can be used to cover a triangle and the user can adjust placement of photos via a control point. However, both tools focus on tile boundary editing and offer limited support for modifying interior designs, making them less effective for effects such as those in Fig.~\ref{fig: lamps} and Fig.~\ref{fig: eye_enlarging}.
\paragraph{Interactive Deformation in Computer Animation:} Since Escher tiles often take the form of animal shapes, to deform them, it is natural to draw inspiration from point handle-based deformation algorithms commonly used in computer animation. These algorithms are designed to produce deformations that are geometrically and physically plausible. For instance, ARAP~\cite{sorkine2007rigid} minimizes a local rigidity energy by computing an optimal rotation field and adjust vertex positions, preserving local shape details during deformation. Bounded biharmonic weights~\cite{jacobson2011bounded} linearly blend the affine transformations of multiple handles to determine the transformation of each point on the shape. The blending weights are computed by minimizing the Laplacian energy, ensuring smooth and natural deformations. However, both methods are iterative approaches based on numerical solving and rely on mesh discretization. Regularized Kelvinlets~\cite{de2017regularized} introduces a mesh-free deformation method by providing a closed-form analytical solution to the elastic wave equation, which models the resulting displacement field. Motivated by these premises, we derive a closed-form solution that periodically deforms tiles while preserving symmetry.
\paragraph{Cone Manifolds and Lattices:} Our method is closely related to and inspired by the concept of Bravais lattice~\cite{smith2002bravais} and the idea of orbifolds (flat cone-manifolds)~\cite{thurston1978geometry, conway2016symmetries}. Orbifolds have previously been employed in computer graphics for parameterization tasks. In~\cite{aigerman2015orbifold}, orbifolds were studied in conjunction with Tutte embeddings, where it was shown that if a tile obtained from cut-open topological sphere, by repeating itself, can seamlessly cover the Euclidean plane without gaps or overlaps, the resulting parameterization is injective. This idea was subsequently extended to non-Euclidean domains, including the hyperbolic plane~\cite{aigerman2016hyperbolic} and the sphere~\cite{aigerman2017spherical}. More closely related to our work is the usage of orbifolds in global parameterization, particularly in efforts to compute symmetric rotational fields~\cite{ray2006periodic, palacios2007rotational}. In practice, however, tiles are only by-products used as a means of proof in these parametrization studies, whereas our work focus on the interactive design of tiles as a format of decorative art. The parametrization works primarily focus on rotational fields, whereas our approach needs to consider all the symmetry operations that appeared in the 17 wallpaper groups. This includes operations such as reflection and glide reflection. The resulted symmetric fields therefore cannot be thoroughly described with special orthogonal groups. Translation is special in the study of tiling in the sense that it's hard to avoid. The scope of our paper does not include aperiodic tiling~\cite{smith2023aperiodic} (the rare scenario where translation is not present). Bravais lattices (also known as translational covers~\cite{10.1145/3005358.3005368, roy2017interactive, zheng2023interactive}) describe pure translation of wallpaper groups with five categories. We use four categories instead, where the centered rectangle lattice was replaced by parallelogram lattices of two directions. In this work, by decomposing the transformations of periodic tiling, we present a new approach that solves the periodic displacement field via closed-form analytical solution.

\section{Background}
\label{chap:background}
Not all tiles or patterns exhibit symmetry. We refer readers to ~\cite{grunbaum1987tilings} for an in-depth review of tiles. In this paper, we focus exclusively on symmetric patterns in the Euclidean plane, which can be fully characterized by the 17 wallpaper groups. Before presenting the closed-form solution, we provide a brief review of the fundamental concepts of symmetry groups relevant to our derivation. For a more comprehensive discussion on symmetries, we refer readers to ~\cite{conway2016symmetries}.

\paragraph{Cone Point:} There exist only two types of cone points on a symmetric pattern. A reflective cone point occurs where two or more reflection lines intersect, with the order indicating how many reflection lines meet at that point. 
 A gyrational cone point serves as a center of rotation, with the order representing the number of rotations required for a point to return to its original position. On the Euclidean plane, cone points can have orders of 2, 3, 4, or 6. 

\paragraph{Wallpaper Group:} There are 17 types of symmetries for planar patterns, known in mathematics as the wallpaper groups. Among the various notation systems for wallpaper groups, we adopt the orbifold signature~\cite{conway2016symmetries}, which provides an intuitive naming convention based on cone points. As such, the 17 wallpaper groups are respectively $\ast 632$, $632$, $\ast 442$, $4\ast2$, $442$, $\ast 333$, $3\ast3$, $333$, $\ast 2222$, $2\ast22$, $22\ast$, $22\times$, $2222$, $o$, $\ast\ast$, $\ast\times$, and $\times\times$. In this notation, "$\ast$" signifies the presence of mirror lines, "$o$" denotes pure translation, "$\times$" denotes glide reflection. Glide reflection acts like a mirror reflection, but there is no mirror line that directly maps a left-facing shape to a right-facing one without first translating along the direction of the mirror line. Digits preceding "$\ast$" indicate gyrational cone points and their corresponding orders, whereas digits following "$\ast$" represent reflective cone points and their orders.

\paragraph{Bravais Lattice:} If a pattern exhibits periodicity, there exists a fundamental unit, identifiable within the pattern, that can be used to tile the entire plane solely through translations. Such translations exhibit a geometric structure, which can take the form of a square (tetragonal lattice, describing groups $\ast442$, $4\ast2$, $442$), rectangle (orthorhombic lattice, describing groups $\ast 2222$, $22\ast$, $\ast\ast$, $\times\times$, $22\times$), parallelogram (monoclinic lattice, describing groups $2222$, $o$, $2\ast22$, $\ast\times$), or hexagon (hexagonal lattice, describing groups $333$, $\ast333$, $3\ast3$, $632$, $\ast632$). Bravais lattice is critical in determining the closed-form solutions, as it specifies the locations where the guidance field lives in the displacement field. The crux of our method is to find Bravias lattice for each wallpaper group and then use it to provide analytical solution.

\section{Method}

\begin{figure}
\includegraphics[width=0.5\textwidth]{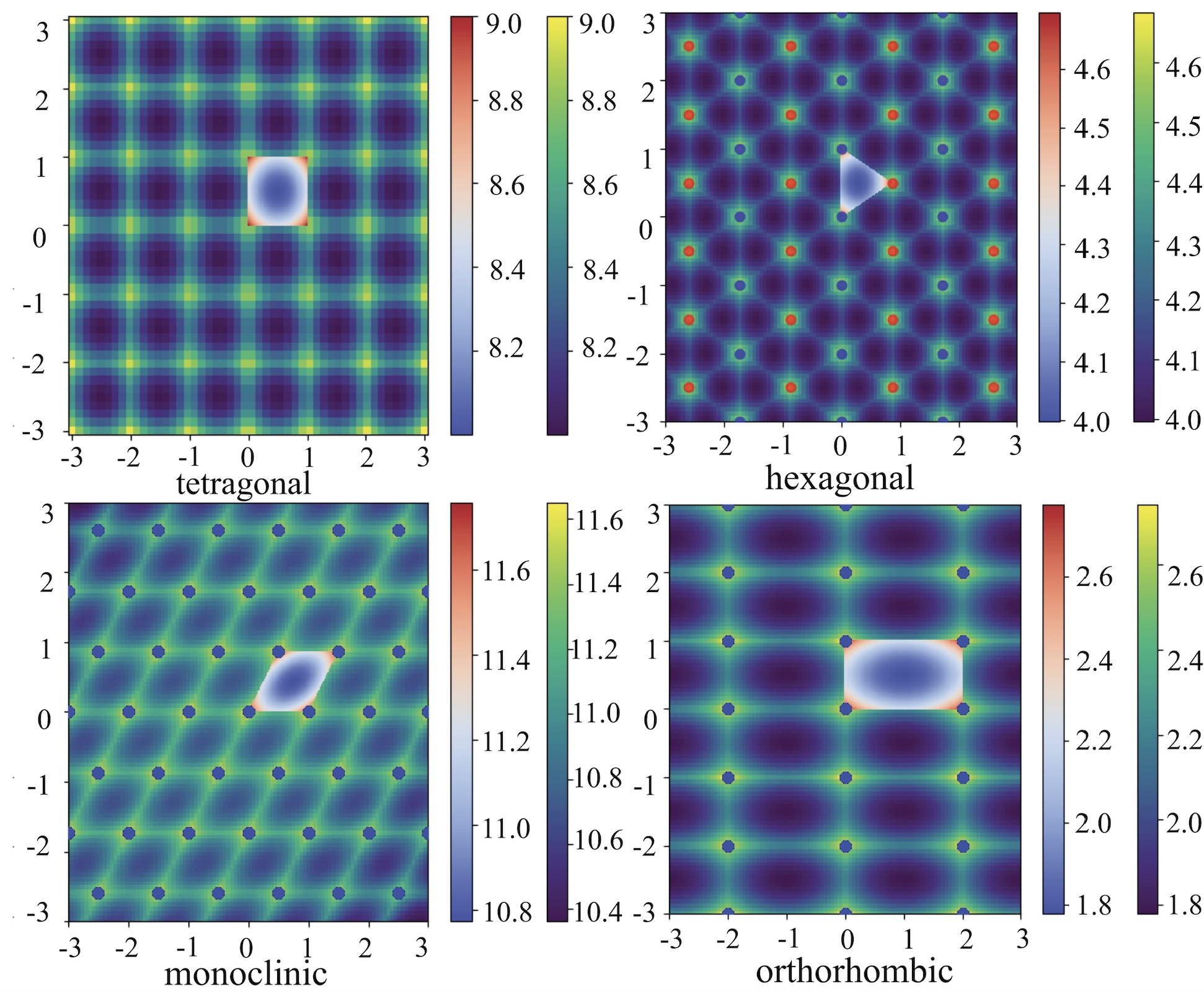}
  \caption{\textbf{Comparison:} We validate the accuracy of our closed-form solution by comparing it with numerical summation. The coolwarm heatmap visualizes the analytical solution within a single cycle, accounting for infinitely many handle copies, while the viridis heatmap represents the numerical solution by summing the contributions of a large number of handles.}
  \label{fig: analytical_numerical_compare}
\end{figure}

When deforming an infinite 2D continuum, the resulting state is determined by a displacement field $\mathbf{u} : \mathbb{R}^2 \to \mathbb{R}^2$. We define a periodic deformation problem that deforms a tessellation with symmetry constraints. The method takes as input a tessellation in various forms, including a 2D triangle mesh or a 2D image mask of a single unit, or an image of the entire pattern. The output is a deformed tessellation, preserving the input's representation while ensuring tileability, seamlessly covering the Euclidean plane without gaps or overlaps. Observing that the displacement field itself is also periodic (like a ``infinite clock", as shown in Fig.~\ref{fig: infinite clock}) $\mathbf{u}(g(x)) = g(\mathbf{u}(x))$ for any symmetry operation $g$, we derive closed-form analytical solutions to enable real-time interactive editing of the tiles. The derivation leverages symbolic computation in Wolfram Mathematica. We further verify the derived real-time closed-form solutions (doing infinite sum of the entire Euclidean plane) by comparing them side by side with slower numerical solutions (doing summation of a large neighborhood) with Fig.~\ref{fig: analytical_numerical_compare}.
Periodic patterns and tilings on the Euclidean plane can be viewed as seamless repeating of orbifolds (that is, disk-like shapes with cone points, which can be folded and joined to form 3D shapes like spheres, tori, or Möbius bands). Our goal is to deform tiles while preserving symmetry constraints. The idea is to leverage an essential property of the orbifold: its symmetry type is determined by the cone points and remains unaffected by the specific choice of cutting paths. The deformation method can be realized by four steps:
\renewcommand{\labelenumi}{\theenumi.}
\begin{enumerate}
    \item Position cone points and mirror lines based on user input.
    \item Place points at the vertices of a cell of the Bravais lattice based on user input.
    \item Position the point handle at the control point $v$ and associate a displacement vector $u_{0}$ according to user mouse drag.
    \item Deform the tiles by integrating a periodic vector field with closed-form solution.
\end{enumerate}

Due to the repeating nature of the tessellation, vector $u_0$ can replicate into infinite copies. Some of these copies may differ from the original vector because the tessellation's symmetry group can alter the vector's direction. We use cone points from Step 1. to calculate the set of distinct vectors we call fundamental vectors \( \{ u_i \mid i \in \mathbb{Z}^+ \cup \{0\}, \; i \leq k \} \) generated from $u_0$. To deform the tessellated pattern, we seek to calculate a displacement vector $u_{\mathbf{x}}$ for each given point $\mathbf{x} \in \mathbb{R}^2$ by composing the effect of all vectors in the set
\begin{equation}
u_{\mathbf{x}} = \sum_{i=0}^k w_{\mathbf{x}i} \cdot u_{i}.
\label{eq:numerical_sum_442}
\end{equation}
The main challenge is calculating the weight $w_{\mathbf{x}i}$. Since the closed-form solution for each fundamental vector has the same color map and differs only by translation, it needs to be computed only once. To simplify the notation in the later derivation, we use $w_{\mathbf{x}}$ instead of $w_{\mathbf{x}i}$.

\begin{figure*}[htb]
\includegraphics[width=1.0\textwidth]{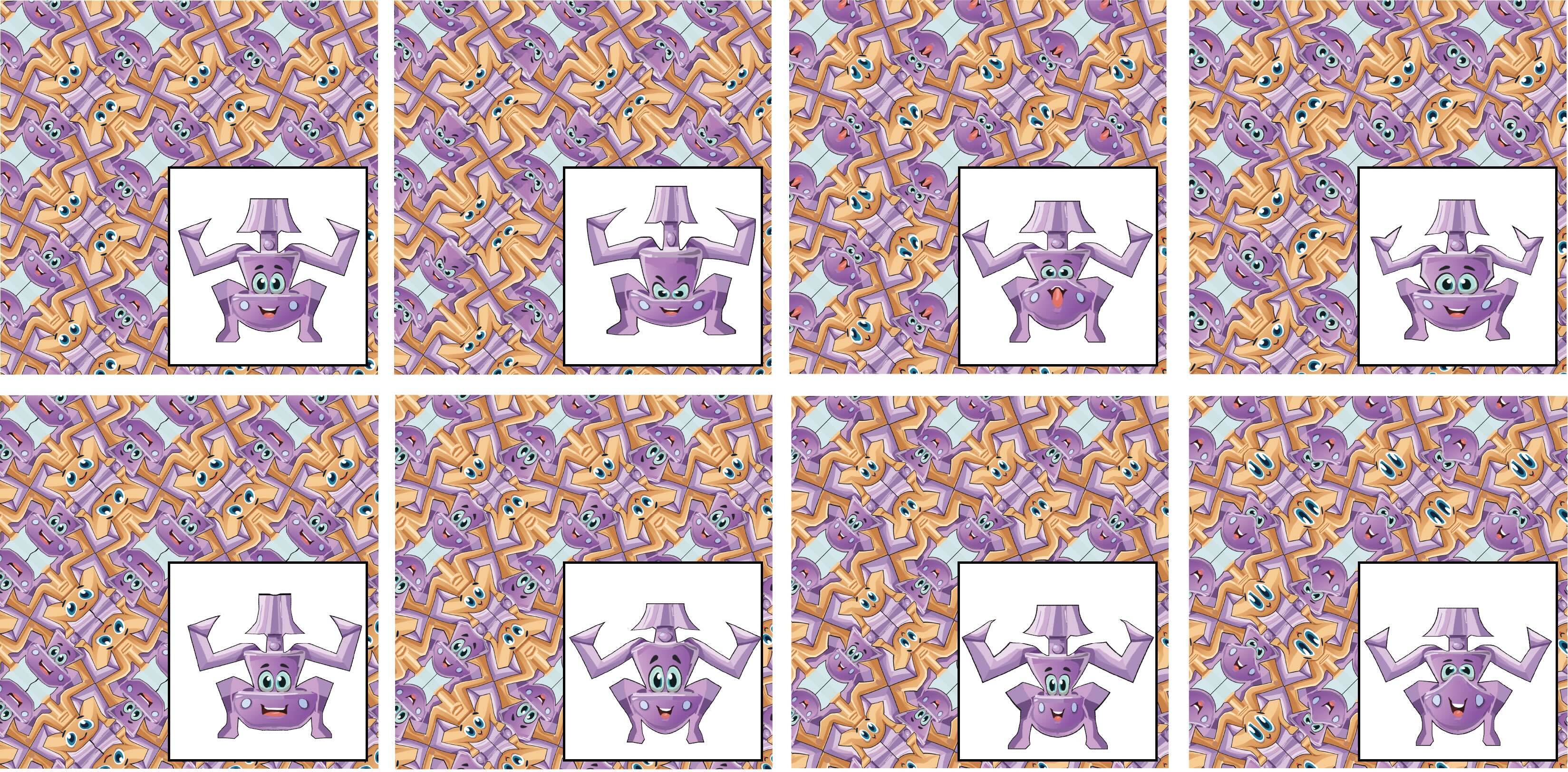}
  \caption{\textbf{Deformable tiles for animation:} Artists can create diverse facial expressions while keeping the lamp a valid tile.}
  \label{fig: purple_yellow_lamp}
\end{figure*}
\subsection{Tetragonal Lattice}
A geometric structure exhibiting the pure translation symmetry of a wallpaper group always exists. The simplest example of such a structure is the tetragonal lattice, consisting of a regular grid of squares. For these symmetry groups, the three cone points locate at the vertices of an isosceles right triangle. The cone points, along with the mirror lines, define a set of transformations that generate the fundamental vectors from a single user-input vector. The cone points and mirror lines automatically determine the presence of eight fundamental vectors. The arrows in each visual depict the guidance field generated by a single fundamental vector.\begin{wrapfigure}{r}{0.25\textwidth}
  \centering
  \includegraphics[width=\linewidth]{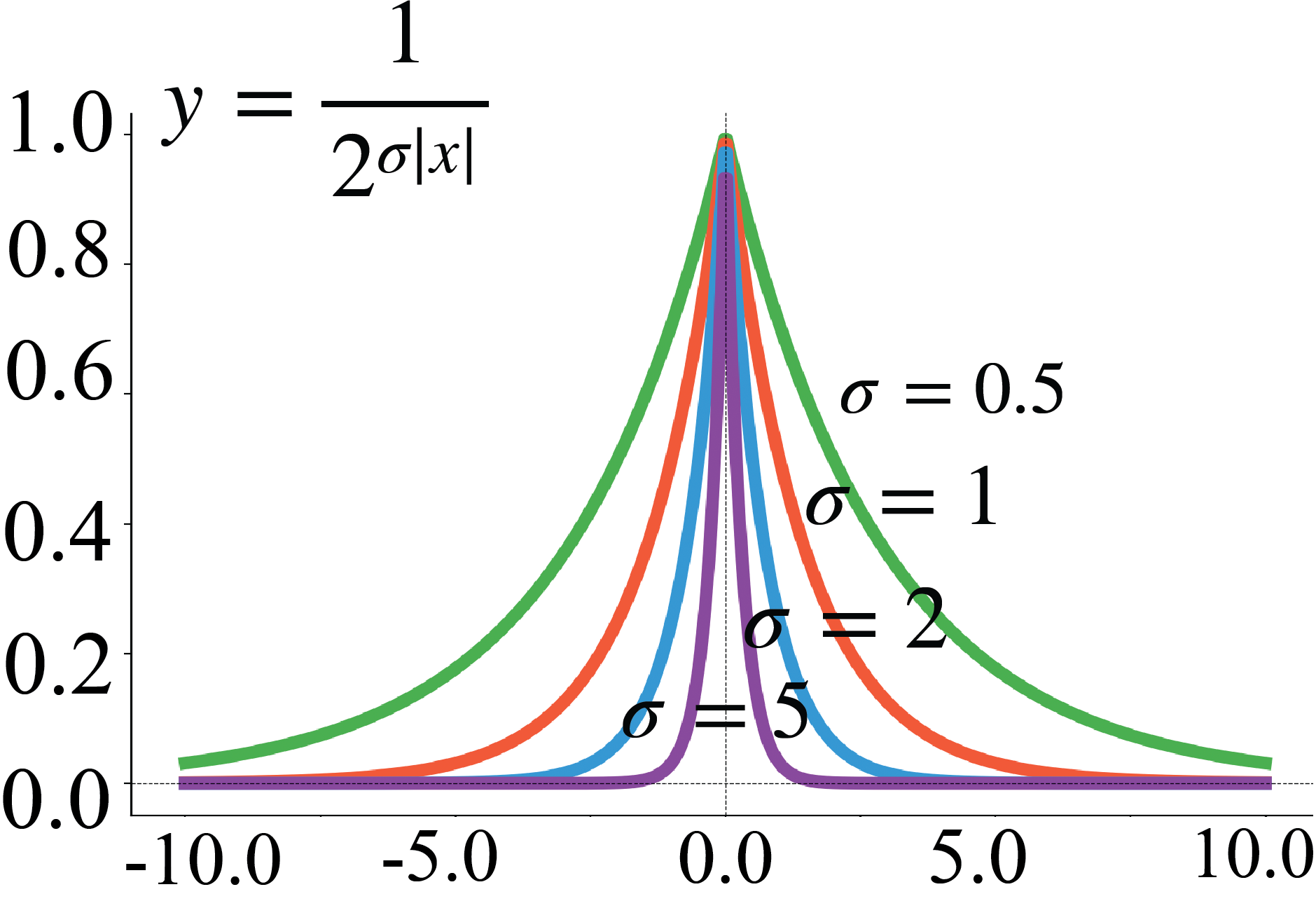}
\end{wrapfigure}

 We model the user's push or pull at a single point $v$ as a translation, represented by a displacement vector $u_0$. Then $u_0$ would act like a force load that changes the shape of the tile. It is desirable that the deformation is local and smooth. Therefore, a predefined falloff function accompanies $v$. We associate each control point $v = (m,n)$ with an auxiliary falloff function
\begin{equation}
c = \frac{1}{2^{\sigma |x-n| + \sigma |y-m|}},
\label{eq:auxiliary function}
\end{equation}
which defines a gradual falloff extending from the handle to distant locations.  A smaller $\sigma$ results in smoother deformations that affect a longer range, whereas a larger $\sigma$ leads to more localized and sharper deformations. Our selection of the falloff function guarantees its infinite sum is the geometric series, which not only converge but also allow for analytical computation, ensuring the deformation algorithm runs in real-time.  This enables us to derive an analytical solution when summing the effects using an infinite series. It is worth noting that only a small fraction of mathematical series converge, and an even smaller subset permits analytical computation. If we select $c = \frac{1}{r}$, where $r$ represents the distance to the load, the summation would be a harmonic series, which does not converge. If we select $c = A \cdot e^{-\frac{(x - n)^2}{2\sigma_x^2} - \frac{(y - m)^2}{2\sigma_y^2}}$, the summation while converge does not have a well-known analytical solution. 
It is not entirely surprising that an analytical solution for tile manipulation can be derived using mathematical series, as both tiles and mathematical series address the challenge of representing infinity within a discrete domain. Mathematical series inherently sum infinitely many terms at fixed intervals, making them a powerful tool for modeling the point handles on a tessellation.
\begin{wrapfigure}{r}{0.15\textwidth}
  \centering
\includegraphics[width=\linewidth]{figures/tetragonal.png}
\end{wrapfigure}

As the tile repeats infinitely, each fundamental vector recurs at the vertices of the lattice grid, forming a guidance field.
\begin{figure}
\includegraphics[width=0.5\textwidth]{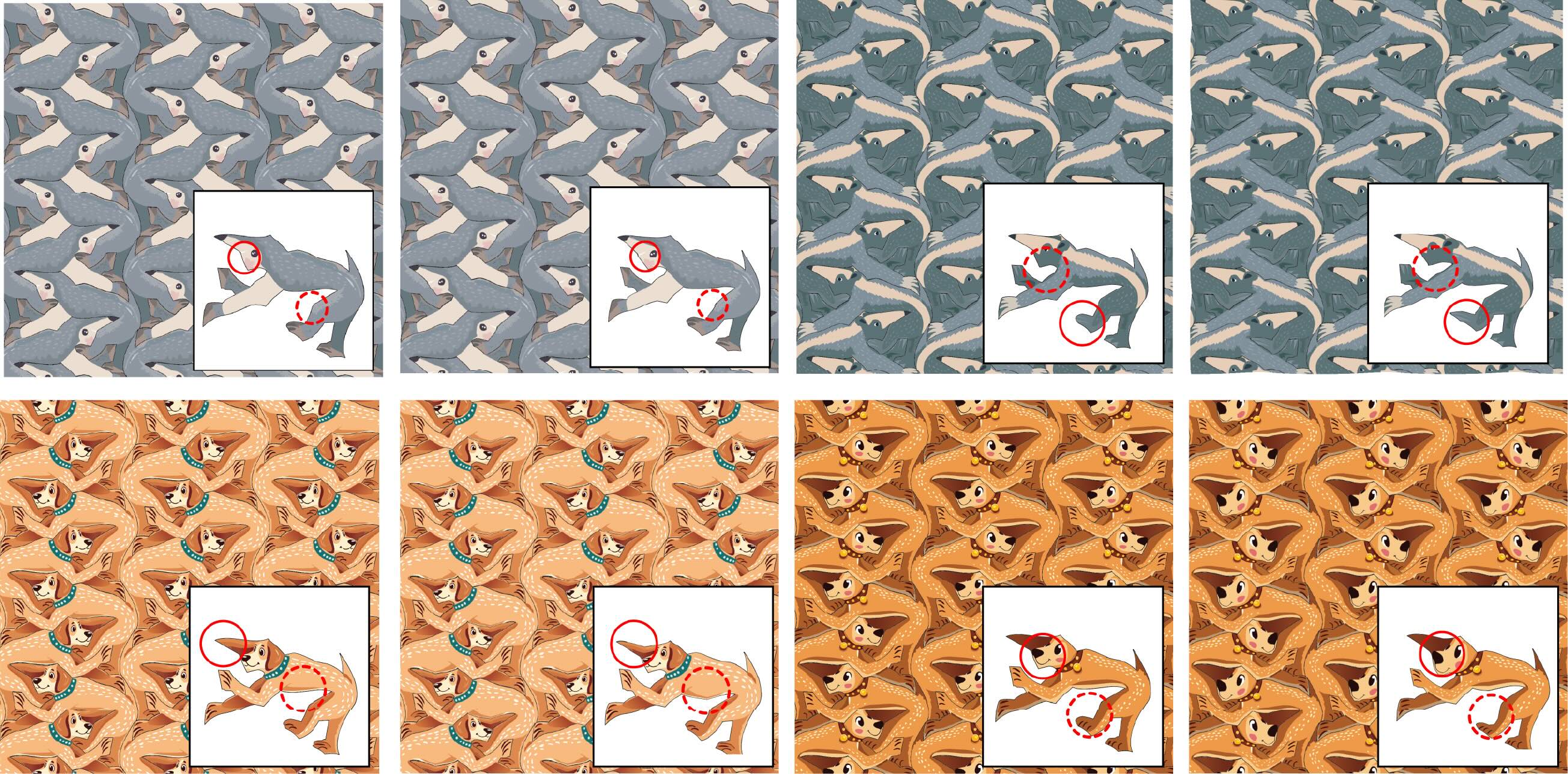}
  \caption{\textbf{Deforming tiles of orthorhombic lattice:} Deformation of dog tile of symmetry group $\times\times$. The solid circles specify the deformation near the user-input handle. The dashed circles specify the deformation caused by the infinite clock so that the resulted shape remains a valid tile, covering the plane with no gap and no overlap.}
  \label{fig: dogs}
\end{figure}
We integrate the effect from vectors living on the regular grid \( \{ (m, n) \mid m, n \in \mathbb{Z} \} \) to calculate the weight
\begin{equation}
w_{\textbf{x}} = \sum_{n = -\infty}^{\infty}\sum_{m = -\infty}^{\infty}\frac{1}{2^{\sigma |x-n| + \sigma |y-m|}}.
\label{eq:numerical_sum_442}
\end{equation}
Due to the periodicity, the closed-form solution only needs to be determined within a single cycle—specifically, one lattice cell, as illustrated by the green square. To query the deformation of arbitrary point on $\mathbb{R}^2$, we can use the square lattice identified in Step 2. to convert the global coordinate $(x_g, y_g)$ of each point $\textbf{x}$ into local coordinate within a unit square cell 
\begin{equation}
(x, y) = \frac{\left| \big( (x_g, y_g) - (x_0, y_0) \big) \mod L \right|}{L},
\label{eq:numerical_sum_442}
\end{equation}
where \( L \) is the side length of the square, \( x_0, y_0 \) is the coordinate of control point $v$,  \( x_g, y_g \) is the global coordinate of a point on the tessellated pattern. The coordinate transformation ensures that identical features repeating at different locations in the global coordinate system share the same local coordinates. This aligns with the orbifold concept, where equivalent features are counted only once through a process of ``folding."~\cite{conway2016symmetries}

Let us focus on getting the closed-form solution of $w_{\textbf{x}}$ in one cell, $x, y \in (0, 1)$. Since Mathematica cannot process mathematical series involving absolute value signs, we decompose the calculation of $w$ into four separate terms to eliminate them. Specifically, the ranges are: $n \geq 1$, $m \geq 1$ for $w_{++}$; $n \leq 0$, $m \leq 0$ for $w_{--}$; $n \geq 1$, $m \leq 0$ for $w_{+-}$; and $n \leq 0$, $m \geq 1$ for $w_{-+}$.

Summing the geometric series, we have the analytical solution
\begin{equation}
\boxed{
\begin{aligned}
w_{\textbf{x}} &= \frac{2^{\sigma (1-x) + \sigma (1-y)}}{(2^{\sigma}-1)^2} 
+ \frac{2^{\sigma x + \sigma y}}{(2^{\sigma}-1)^2} \\
&+ \frac{2^{\sigma x + \sigma (1-y)}}{(2^{\sigma}-1)^2} 
+ \frac{2^{\sigma (1-x) + \sigma y}}{(2^{\sigma}-1)^2}.
\end{aligned}
}
\label{closed_form_solution_orbifold_442}
\end{equation}
Applying the solution to different user-defined handles produces the deformations shown in Fig.~\ref{fig: purple_yellow_lamp}.

\subsection{Orthorhombic Lattice} 
A natural generalization from the tetragonal lattice is the orthorhombic lattice, featuring a grid of rectangles. We can easily convert the global coordinate into local coordinate within the range of $(0, 1)$ by
\begin{equation}
\left\{
\begin{aligned}
x &= \frac{|x_g + (L_x - x_0)| \mod L_x}{L_x} \\
y &= \frac{|y_g + (L_y - y_0)| \mod L_y}{L_y}.
\end{aligned}
\right.
\end{equation}
Then the rest could leverage Eq.~\ref{closed_form_solution_orbifold_442}.

Figs.~\ref{fig: dogs}  shows examples of deforming dog and anteater tiles with orthorhombic lattice. 
\begin{wrapfigure}{r}{0.15\textwidth}
  \centering
\includegraphics[width=\linewidth]{figures/hexagonal.png}
\end{wrapfigure}
\subsection{Hexagonal Lattice}
Another type of symmetry features a hexagonal lattice, which consists of a regular grid of equilateral triangles. The three cone points lie on the vertices of a 30-60-90 triangle or an equilateral triangle. 

We associate each control point \( \{ (n, \sqrt{3}m) \mid m, n \in \mathbb{Z} \} \) with an auxiliary function
\begin{equation}
c_1 = \frac{1}{2^{\sigma |\frac{\sqrt{3}x+y-\sqrt{3}n-\sqrt{3}m}{2}| + \sigma |\frac{\sqrt{3}x-y-\sqrt{3}n+\sqrt{3}m}{2}| + \sigma |-y+\sqrt{3}m|}},
\label{c1}
\end{equation}
and associate with each control point \( \{ (\frac{2n+1}{2}, \frac{2\sqrt{3}m+\sqrt{3}}{2}) \mid m, n \in \mathbb{Z} \} \) a slightly different auxiliary function
\begin{equation}
c_2 = \frac{1}{2^{\sigma|\frac{\sqrt{3}x+y-\sqrt{3}n-\sqrt{3}m-\sqrt{3}}{2}|+\sigma|\frac{\sqrt{3}x-y-\sqrt{3}n+\sqrt{3}m}{2}|+\sigma|-y+\sqrt{3}m+\frac{\sqrt{3}}{2}|}}.
\label{c2}
\end{equation}
This arrangement arises from the auxiliary function being smooth except along three ridges.  Aligning these ridges with the edges of the lattice grid ensures the smoothness of the resulting deformation.

\begin{figure}
\includegraphics[width=0.47\textwidth]{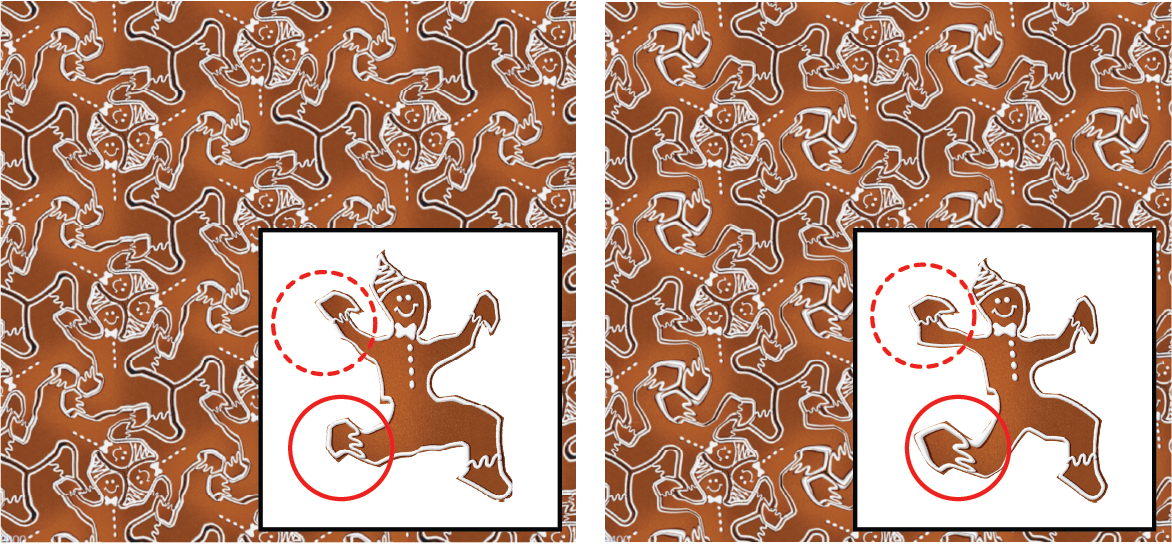}
  \caption{\textbf{Deforming tiles of hexagonal lattice:} Deformation of gingerbread tile of symmetry group 333. This example shows that the user wants to edit the leg of the gingerbread so that it looks like he is running. Because the puzzle is like an ``infinite clock", when the leg deforms, an arm deforms with it for the shape to remain a valid tile.}
  \label{fig: ginger}
\end{figure}

The one cycle that the closed-form solution should live in would be an equilateral triangle as illustrated by the green triangle where
\begin{equation}
\left\{
\begin{aligned}
&0 < y < \frac{\sqrt{3}}{2} \\
&\frac{\sqrt{3}}{3}y< x < -\frac{\sqrt{3}}{3}y + 1\\
\end{aligned}
\right.
\end{equation}
For a point $\textbf{x}$ in the picked equilateral triangle cell, we pursue the closed-form solution
\begin{equation}
w_{\textbf{x}} = \sum_{n=-\infty}^{\infty}\sum_{m=-\infty}^{\infty}c_1(x,y) + \sum_{n=-\infty}^{\infty}\sum_{m=-\infty}^{\infty}c_2(x,y).
\label{two terms}
\end{equation}
We have the closed-form analytical solution of the weight
\begin{equation}
\boxed{
\begin{aligned}
w_{\textbf{x}} &= 
\frac{2^{\sigma(-y - \sqrt{3} (-3 + x))}}{(-1 + 2^{\sigma\sqrt{3}})^2 (1 + 2^{\sigma\sqrt{3}})} 
+ \frac{2^{1 + 2 \sqrt{3} \sigma - 2 y \sigma}}{(-1 + 4^{\sqrt{3}\sigma})^2} \\
&+ \frac{2^{\sigma(-y + \sqrt{3} (2 + x))}}{(-1 + 2^{\sigma\sqrt{3}})^2 (1 + 2^{\sigma\sqrt{3}})} 
+ \frac{2^{\sigma(y - \sqrt{3} (-1 + x))}}{(-1 + 2^{\sigma\sqrt{3}})^2 (1 + 2^{\sigma\sqrt{3}})}\\
&+ \frac{2^{1 + 2 \sqrt{3}\sigma + 2 y \sigma}}{(-1 + 4^{\sqrt{3} \sigma})^2} 
+ \frac{2^{\sigma(y + \sqrt{3} x)}}{(-1 + 2^{\sigma\sqrt{3}})^2 (1 + 2^{\sigma\sqrt{3}})}\\
&+ \frac{2^{\sigma (-y - \sqrt{3} (-2 + x))}}{(-1 + 2^{\sigma\sqrt{3}})^2 (1 + 2^{\sigma\sqrt{3}})} 
+ \frac{2^{\sigma(\sqrt{3} - 2y)} \left(1 + 4^{\sigma\sqrt{3}}\right)}{\left(-1 + 4^{\sigma\sqrt{3}}\right)^2}\\
&+ \frac{2^{\sigma (-y + \sqrt{3}(1 + x))}}{(-1 + 2^{\sigma\sqrt{3}})^2(1 + 2^{\sigma\sqrt{3}})} 
+ \frac{2^{\sigma (y - \sqrt{3} (-2 + x))}}{(-1 + 2^{\sigma \sqrt{3}})^2 (1 + 2^{\sigma \sqrt{3}})}\\
&+ \frac{2^{\sigma(\sqrt{3} + 2y)} (1 + 4^{\sigma\sqrt{3}})}{(-1 + 4^{\sigma\sqrt{3}})^2} 
+ \frac{2^{\sigma (y + \sqrt{3} (1 + x))}}{(-1 + 2^{\sigma\sqrt{3}})^2 (1 + 2^{\sigma\sqrt{3}})}.
\end{aligned}
\label{closed_form_solution_hexagon}
}
\end{equation}
Figs.~\ref{fig: ginger} shows an example of deforming tiles with hexagonal lattice. 
\begin{figure}[h]
\includegraphics[width=0.47\textwidth]{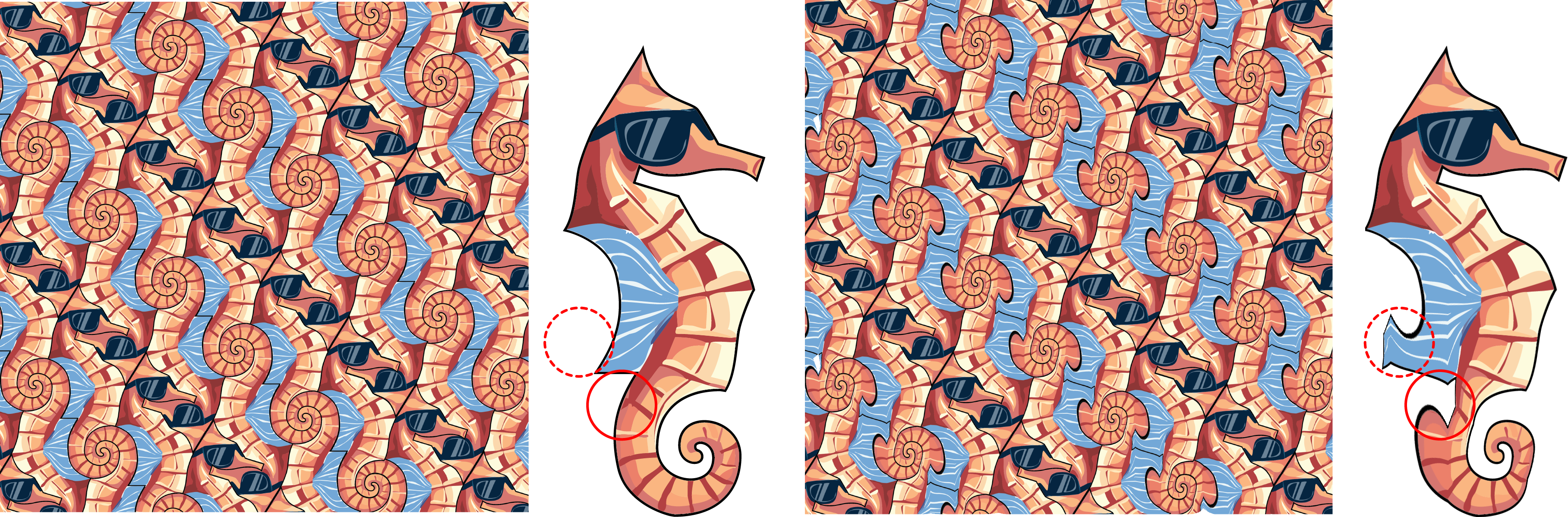}
  \caption{\textbf{Deforming tiles of monoclinic lattice:} Deformation of seahorse tile of symmetry group 2222. This example shows that the user wants to edit the tail of the seahorse. Because the puzzle is like an ``infinite clock", when the tail deforms, the wing deforms with it for the tile to remain valid.}
  \label{fig: seahorse}
\end{figure}
\subsection{Monoclinic Lattice}
The last type of structure is the monoiclinic lattice, feaaturing a grid of parallelograms. The four cone points of $2222$ lie on a parallelogram. Group o involves only pure translation, allowing us to skip step 1. \begin{wrapfigure}{r}{0.15\textwidth}
  \centering
\includegraphics[width=\linewidth]{figures/monoclinic.png}
\end{wrapfigure} In group $2\ast22$, the user defines two perpendicular sets of mirror lines, with the rectangle's center aligning with the gyrational cone point. In group $\ast\times$, the user specifies a mirror line and a glide reflection.

The closed-form analytical solution resides within a single cycle, represented by the parallelogram:
\[(x, y) = s \mathbf{a} + t \mathbf{b}, \quad \text{where } \mathbf{a} = (a_1, 0), \, \mathbf{b} = (b_1, b_2), \, 0 \leq s, t \leq 1.\] Here, \((x, y)\) denotes the Cartesian coordinates, and \((s, t)\) represents the affine coordinates defined by the decomposition of any point \((x, y)\) as a linear combination of \(\mathbf{a}\) and \(\mathbf{b}\). The length of \(\mathbf{a}\) and \(\mathbf{b}\) are respectively \(L_1\) and \(L_2\).

We associate to each control point an auxiliary function
\begin{equation}
c = \frac{1}{2^{\sigma|L_1(t-n)| + \sigma|L_2(s-m)|}}.
\end{equation}

For a point $\textbf{x}$ in the picked parallelogram cell, we pursue the closed-form solution
\begin{equation}
w_{\textbf{x}} = \sum_{n=-\infty}^{\infty}\sum_{m=-\infty}^{\infty} \frac{1}{2^{\sigma|L_1(t-n)| + \sigma|L_2(s-m)|}}.
\end{equation}
We have the closed-form analytical solution of the weight
\begin{equation}
\boxed{
\begin{aligned}
w_{\textbf{x}} &= \frac{2^{L_1 \sigma + L_2 \sigma - L_2 s \sigma - L_1 t \sigma}}
{(-1 + 2^{L_1 \sigma})(-1 + 2^{L_2 \sigma})} + \frac{2^{L_2 s \sigma + L_1 t \sigma}}{(-1 + 2^{L_1 \sigma})(-1 + 2^{L_2 \sigma})} \\
&+ \frac{2^{L_2 \sigma - L_2 s \sigma + L_1 t \sigma}}{(-1 + 2^{L_1 \sigma})(-1 + 2^{L_2 \sigma})} + \frac{2^{L_1 \sigma + L_2 s \sigma - L_1 t \sigma}}{(-1 + 2^{L_1 \sigma})(-1 + 2^{L_2 \sigma})}
\end{aligned}
}
\label{closed_form_solution_parallelogram}
\end{equation}
Figs.~\ref{fig: seahorse} shows an example of deforming the seahorse tile of monoclinic lattice. 

\section{Experiment and Analysis}
\begin{figure}[htb]
\includegraphics[width=0.5\textwidth]{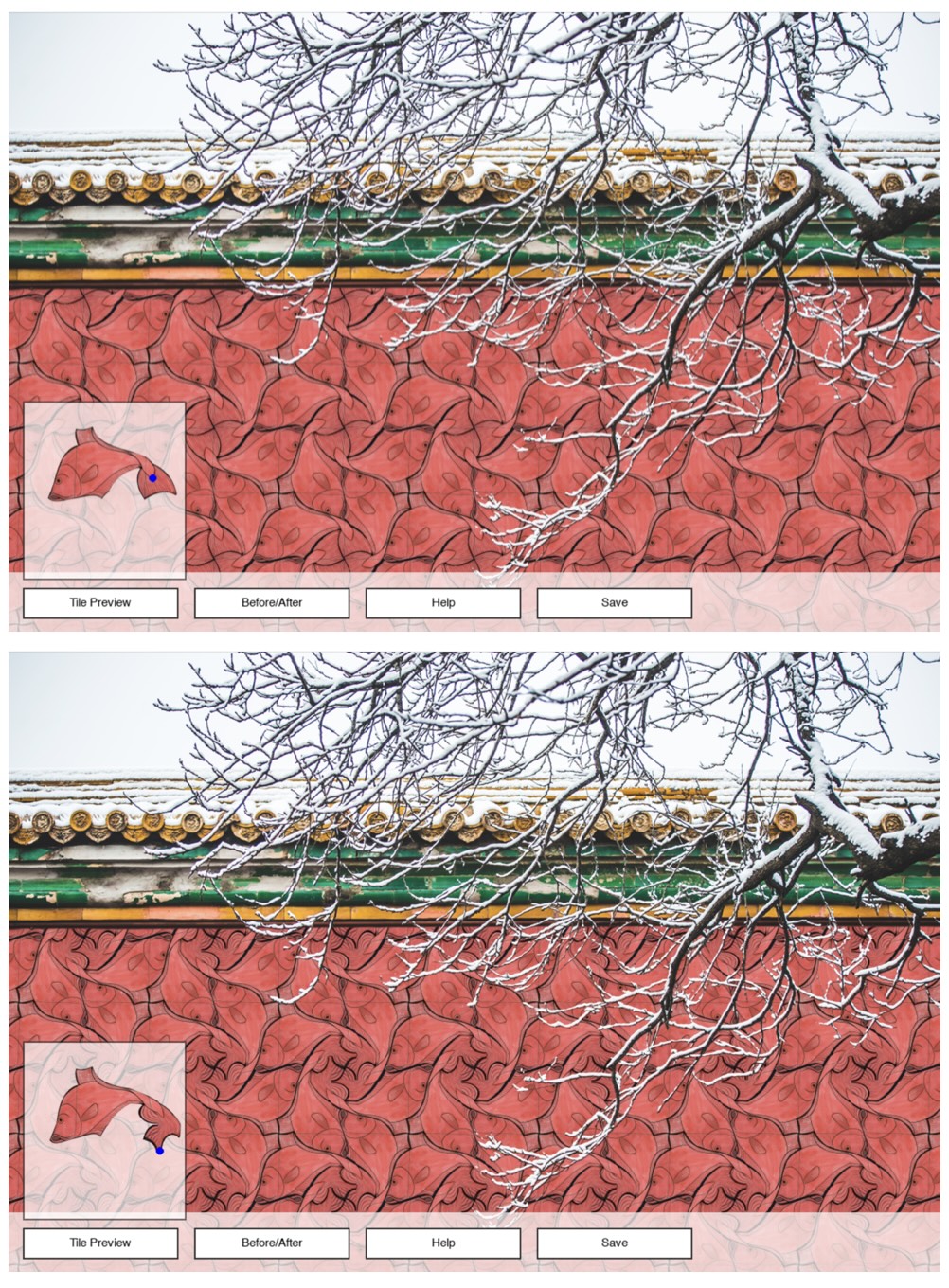}
  \caption{\textbf{User interface of the photo editing tool:} A point handle can be placed on the tile preview window at the lower left corner of the screen. The entire pattern in the photo will deform in real time according to the drag in the preview window. In this example, the user wants to deform the tail of the fish so that it looks like a heart shape. Top: before. Bottom: after. All M.C. Escher works © 2025 The M.C. Escher Company - the Netherlands. All rights reserved. Used by permission. www.mcescher.com }
  \label{fig: user interface}
\end{figure}

\paragraph{Real-Time Tile Deformation:} The user interface of the photo editing tool (Fig.~\ref{fig: user interface}) provides an intuitive platform for interactive tile deformation. Users can place a point handle on the tile preview window and drag it to modify the pattern in real time. The deformation is applied seamlessly across the entire tessellation, maintaining tileability and symmetry constraints. The before and after states of the deformation process are displayed side-by-side, demonstrating the precision and fluidity of the tool.

\begin{figure}[h]
\includegraphics[width=0.5\textwidth]{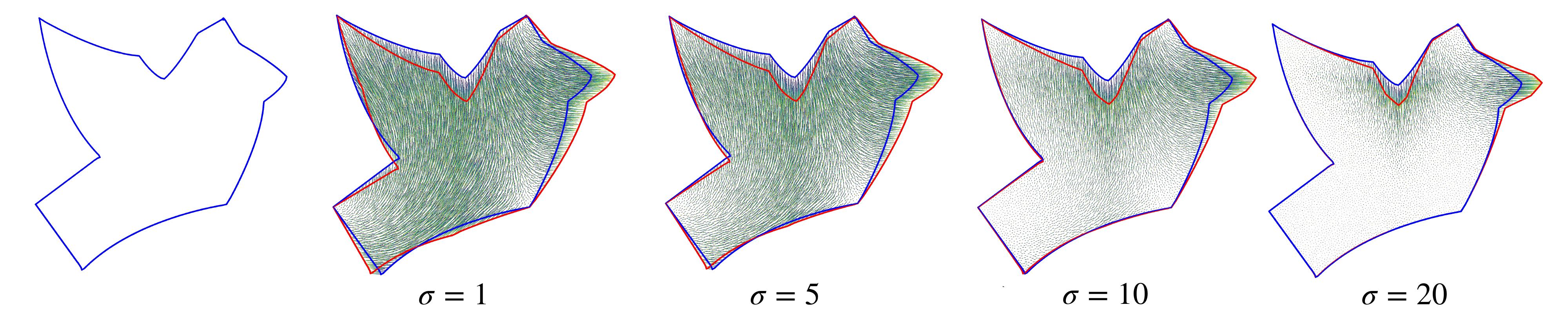}
  \caption{\textbf{Locality control:} Users can alter how local or global they wish for the brush by adjusting $\sigma$. The extruding of beak corresponds to intruding of the back. Larger $\sigma$ corresponds to more localized deformation near the push. Smaller $\sigma$ corresponds to more global deformation which impacts the entire curve.}
  \label{fig: sigma}
\end{figure}

\begin{figure}
\includegraphics[width=0.5\textwidth]{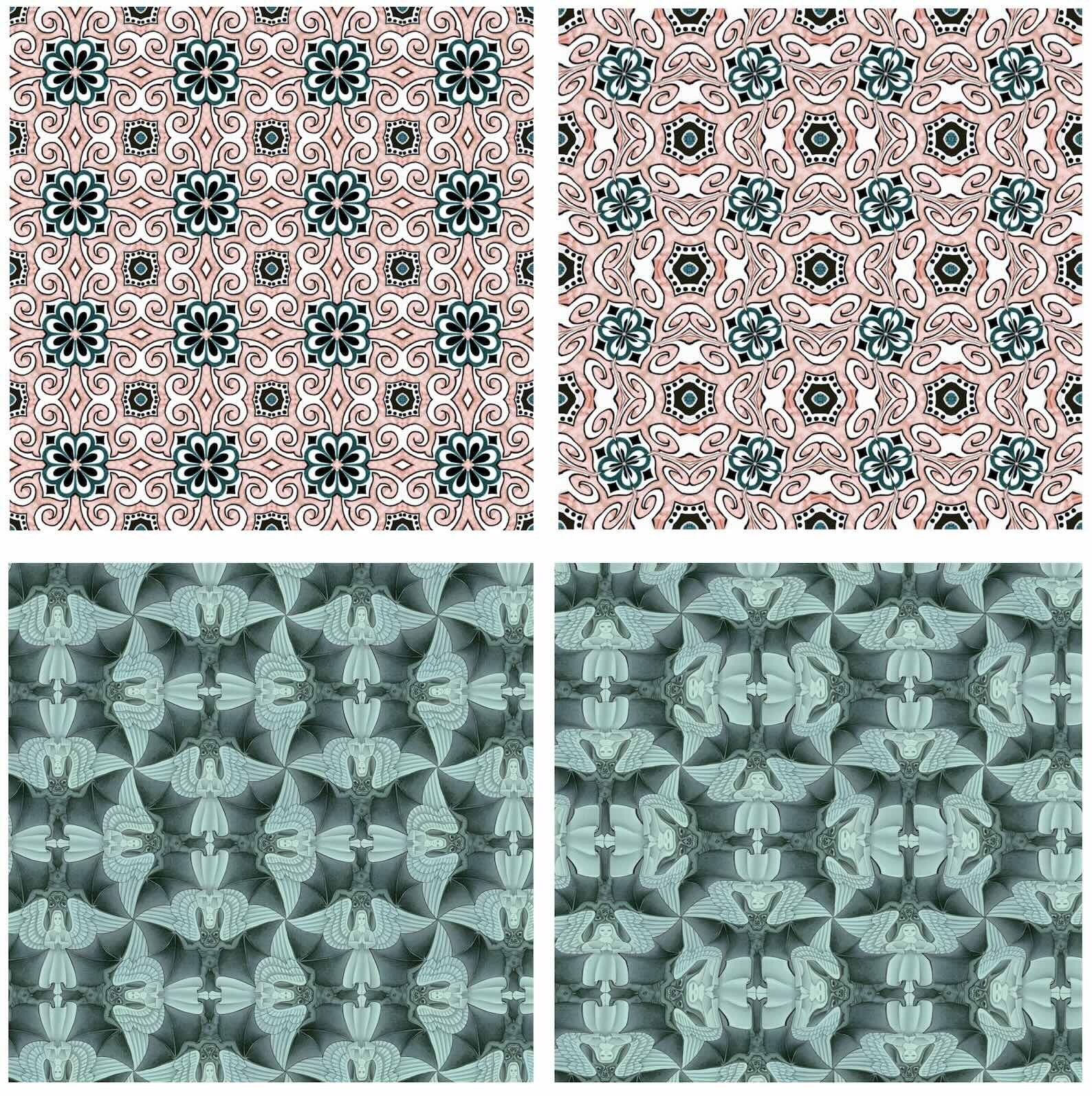}
  \caption{\textbf{Jump of the level of symmetry:} Deformation can happen across different symmetry groups. The first row is feeding a $\ast442$ pattern into a 442 deformer, getting a $4\ast2$ as a result. The second row is feeding a $4\ast2$ pattern into a 2222 deformer, getting a $\ast2222$ pattern as a result. Second row: All M.C. Escher works © 2025 The M.C. Escher Company - the Netherlands. All rights reserved. Used by permission. www.mcescher.com }
  \label{fig: change_symmetry_type}
\end{figure}

\begin{figure}
\includegraphics[width=0.5\textwidth]{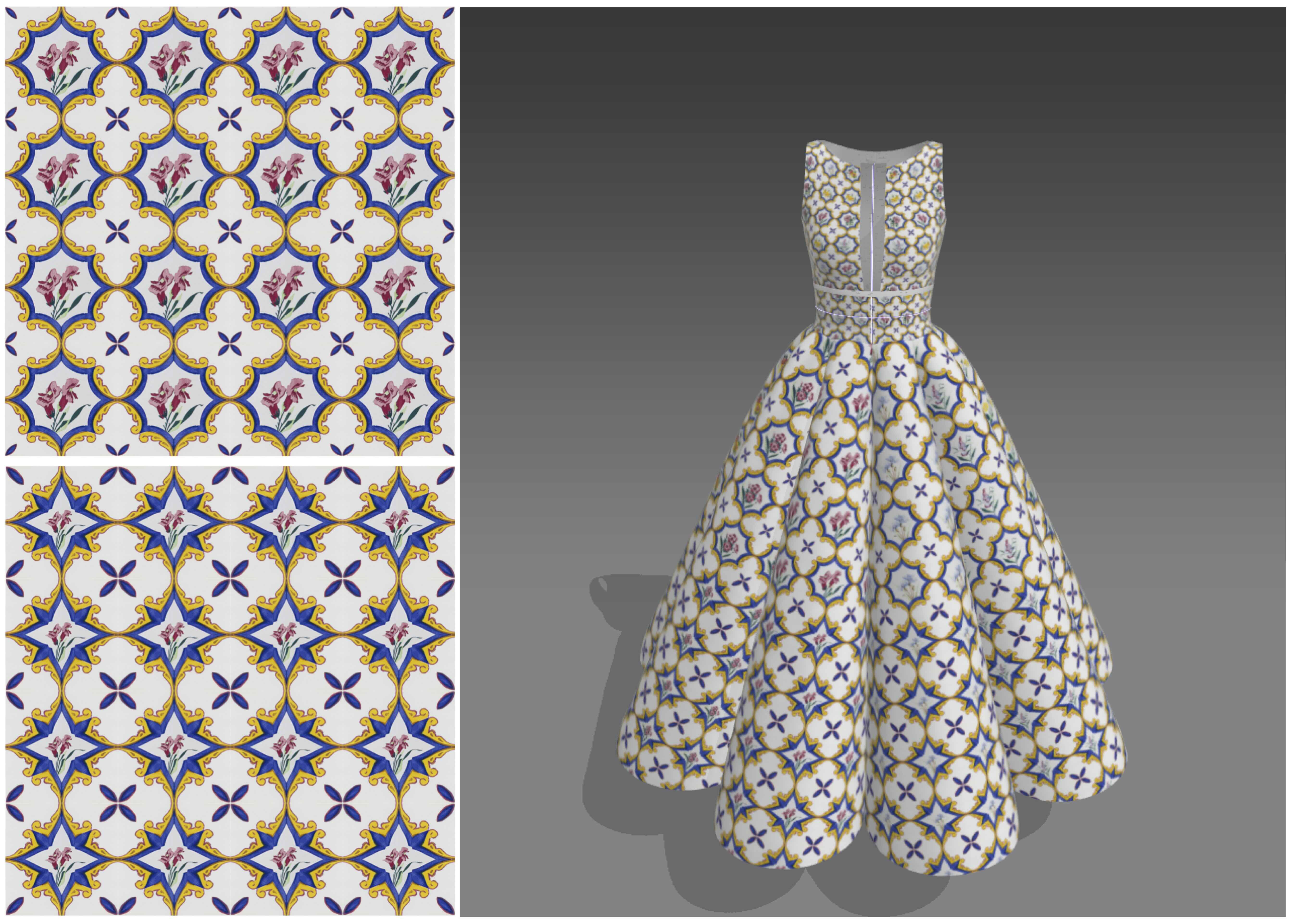}
  \caption{\textbf{Deformable tiles for fabrication:} The pattern is deformed from top left to top bottom. Both before and after patterns are applied in the design of the dress.}
  \label{fig: dress}
\end{figure}

\begin{figure}
\includegraphics[width=0.5\textwidth]{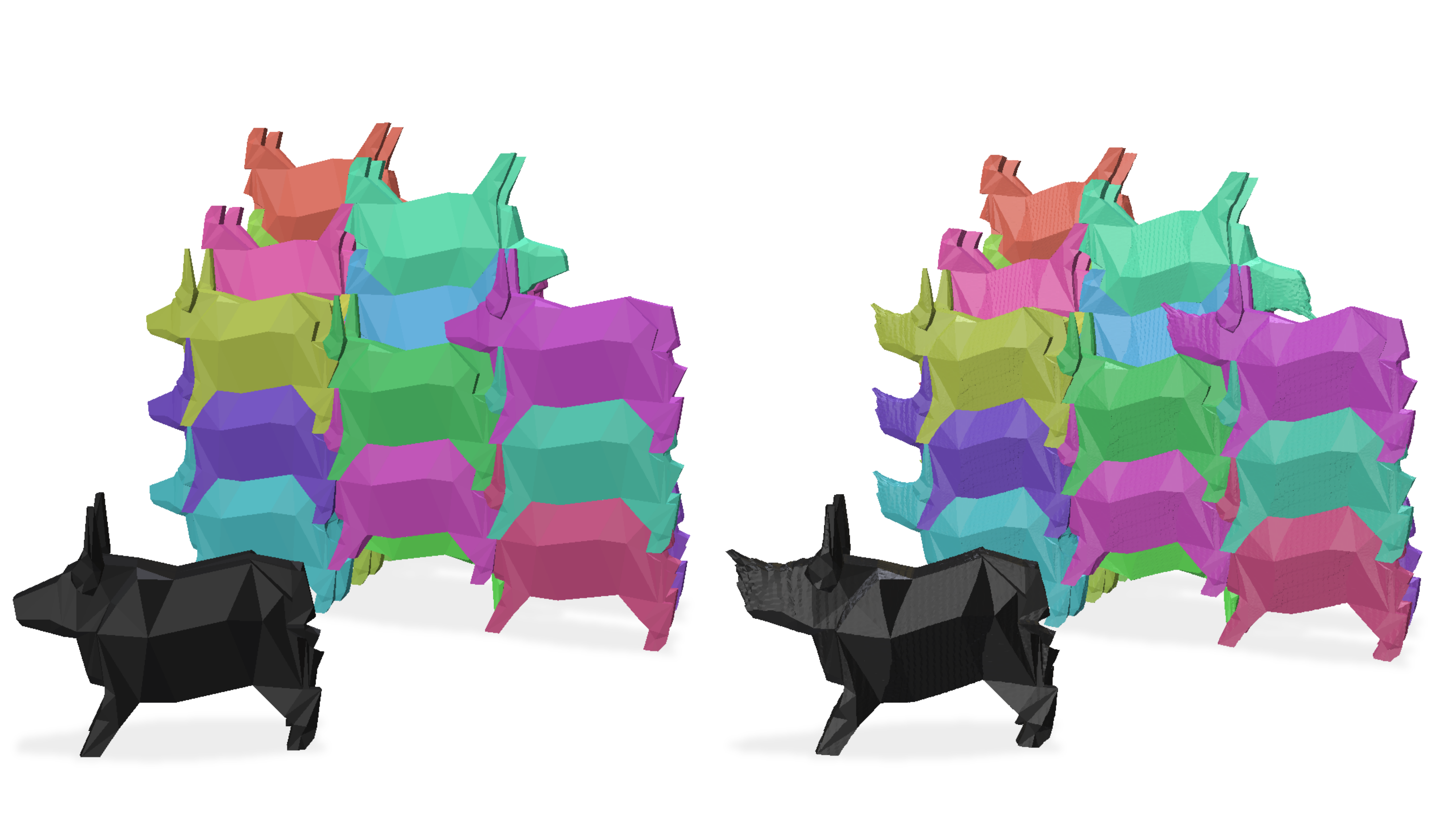}
  \caption{\textbf{Deforming a packable cow into a packable rhino:} This example shows the possibility to sculpt a volume from one animal to another while maintaining the shape as a valid tile.}
  \label{fig: cow_sculpting}
\end{figure}

\paragraph{Locality Control with Adjustable Falloff:} One of the key features of our method is the ability to control the locality of the deformation using the parameter \(\sigma\). As illustrated in Fig.~\ref{fig: sigma}, users can fine-tune \(\sigma\) to adjust the deformation's impact from highly localized edits to more global modifications. This flexibility allows artists to achieve desired effects with minimal effort, ensuring the deformation aligns with the creative intent.
\paragraph{Symmetry Transitions:} Our method supports transformations across symmetry groups by leveraging the hierarchical relationships between them, where subgroups represent a downgrade in the level of symmetry. This capability provides unprecedented flexibility for artistic exploration. Fig.~\ref{fig: change_symmetry_type} illustrates this with two examples. In the first row, a $\ast442$ pattern is transformed into a $4\ast2$ pattern, representing a reduction in symmetry as $4\ast2$ is a subgroup of 442. In the second row, a $4\ast2$ pattern is further deformed into a *2222 pattern, reflecting another symmetry downgrade. 
\paragraph{Applications in Fabrication:} The deformable patterns generated by our method can be directly applied to real-world design scenarios, such as textile fabrication. Fig.~\ref{fig: dress} highlights a dress design incorporating a tile pattern that has been deformed. The results showcase the aesthetic versatility of the tool, enabling seamless integration of complex patterns into wearable designs.
\paragraph{Sculpting and Creative Exploration:} Beyond tiling and fabrication, our method enables creative sculpting of patterns. The results validate the method's ability to maintain periodicity and symmetry while allowing for flexible deformation. Fig.~\ref{fig: cow_sculpting} demonstrates how a "packable cow" tile can be deformed into a "packable rhino" tile. The generalization is achieved by extending one dimension from the 2D parallelogram cells into 3D parallelepiped cells. The falloff function is in this case geometric series with three variables. 
\section{Limitations \& Future Work}
We introduced a deformation method with symmetry constraints. It can be used to deform tessellations without introducing gaps or overlaps.  Our closed-form solution enables simultaneous deformation of tile boundaries and interiors in real-time. An adaptive fall-off parameter allows both localized and global adjustments, enhancing artistic flexibility.
Our work is an early step in tiling-aware deformation. We believe the two most important limitations to tackle in the near future are: (i) ensure injectivity of the map under extreme deformations, and (ii) explore a wider range of symmetry downgrading scenarios. We hope to tackle the first challenge by designing a falloff function that offers divergence-free deformations solving the elastic wave equation ~\cite{de2017regularized, de2018dynamic} via Tutte embedding. A promising direction to explore is to leverage Kelvinlets~\cite{phan1994microstructures} to find a closed-form solution that guarantees injectivity. For the second challenge, a comprehensive treatment of symmetry downgrading would require enumerating the subgroups of each wallpaper group, adopting these as prior assumptions, and verifying that the deformed pattern satisfies the defining conditions of the corresponding subgroup with mathematical proof.

We also would like to explore applying deformation in the parameterized (UV) space, extending deformable tiles to apply to arbitrary surfaces such as polygonal meshes or analytical surfaces: spheres and hyperbolic planes, and designing semantic interlocking shapes. In addition, our work may contribute an additional practical motivation for further exploration into closed-form mathematical series of Gaussian distributions using theta functions~\cite{bagis2015complete} and elliptic functions~\cite{akhiezer1990elements}.

\begin{acks}
We are grateful to Noam Aigerman and Thibault Groueix for their insightful discussions during the early stages of this project. We thank Abhijeet Chowdhury and Taiwei Cui for their help in improving the GUI. We also appreciate Dan B. Goldman and Alex Suter for feedback on the manuscript; Ruben Wiersma for proofreading and communication with the M.C. Escher Company; Stephen Spencer for guidance on ACM proceedings; and Fernando de Goes for discussions on divergence-free displacement fields.
\end{acks}

\clearpage
\bibliographystyle{ACM-Reference-Format}
\bibliography{sample-bibliography}

\clearpage